\documentclass[sigconf,screen]{acmart}
\usepackage[UKenglish]{babel}
\usepackage{csquotes}
\usepackage{xcolor}
\usepackage{xspace}
\usepackage{subfig}
\usepackage{multirow}
\usepackage{booktabs}
\usepackage{balance}
\usepackage{siunitx}
\usepackage{hyperref}
\usepackage[capitalise]{cleveref}

\newcommand\definetool[2]{\newcommand{#1}{{\emph{#2}}\xspace}}
\newcommand\initialism[2]{\newcommand{#1}{{\textsc{#2}}\xspace}}
\definetool{\Scratch}{Scratch}
\definetool{\whisker}{Whisker}
\definetool{\litterbox}{LitterBox}
\definetool{\scratchblocks}{scratchblocks}
\definetool{\gpt}{GPT-4.1}
\definetool{\openai}{OpenAI}
\initialism{\api}{api}
\initialism{\json}{json}
\initialism{\llm}{llm}
\newcommand{\llms}{\textsc{llm}s\xspace}

\newcommand{\interpretation}[2]{
  \vspace{0.4em}
  \noindent
  \colorbox{gray!20}{\parbox{.97\linewidth}{\if #1\empty\else
      \textbf{#1.}
      \fi #2
    }}}

\newcommand{\score}[1]{\mbox{\ensuremath{\sigma #1}}}
\newcommand{\mscore}[1]{\mbox{\ensuremath{\sigma_m #1}}}
\newcommand{\hscore}[1]{\mbox{\ensuremath{\sigma_h #1}}}
\newcommand{\iscore}[1]{\mbox{\ensuremath{\sigma_i #1}}}
\newcommand{\nscore}[1]{\mbox{\ensuremath{\sigma_n #1}}}
\newcommand{\mean}[1]{\mbox{\ensuremath{\varnothing #1}}}
\newcommand{\mmean}[1]{\mbox{\ensuremath{\varnothing_m #1}}}
\newcommand{\hmean}[1]{\mbox{\ensuremath{\varnothing_h #1}}}
\newcommand{\imean}[1]{\mbox{\ensuremath{\varnothing_i #1}}}
\newcommand{\nmean}[1]{\mbox{\ensuremath{\varnothing_n #1}}}

\newcommand{\fkappa}[1]{\ensuremath{\kappa = \num[round-mode=figures,round-precision=3]{#1}}}

\usepackage[colorinlistoftodos,prependcaption,textsize=tiny]{todonotes}

\begin{document}

\title{Detecting Gender Stereotypes in Scratch Programming Tutorials}

\copyrightyear{2025}
\acmYear{2025}
\setcopyright{cc}
\setcctype{by}
\acmConference[Koli Calling '25]{25th Koli Calling International Conference on Computing Education Research}{November 11--16, 2025}{Koli, Finland}
\acmBooktitle{25th Koli Calling International Conference on Computing Education Research (Koli Calling '25), November 11--16, 2025, Koli, Finland}
\acmDOI{10.1145/3769994.3770019}
\acmISBN{979-8-4007-1599-0/25/11}

\author{Isabella Graßl}
\authornote{Both authors contributed equally to this research.}
\orcid{0000-0001-5522-7737}
\affiliation{\institution{Technical University of Darmstadt}
  \city{Darmstadt}
  \country{Germany}
}
\email{isabella.grassl@tu-darmstadt.de}
\author{Benedikt Fein}
\authornotemark[1]
\orcid{0000-0002-3798-845X}
\affiliation{\institution{University of Passau}
  \city{Passau}
  \country{Germany}
}
\email{benedikt.fein@uni-passau.de}
\author{Gordon Fraser}
\orcid{0000-0002-4364-6595}
\affiliation{\institution{University of Passau}
  \city{Passau}
  \country{Germany}
}
\email{gordon.fraser@uni-passau.de}

\begin{abstract}
Gender stereotypes in introductory programming courses often go
  unnoticed, yet they can negatively influence young learners'
  interest and learning, particularly underrepresented groups such as girls.
Popular tutorials on block-based programming with
  \Scratch may unintentionally reinforce biases through character
  choices, narrative framing, or activity types.
Educators currently lack support in identifying and addressing such
  bias. With large language models~(\llms) increasingly used to
  generate teaching materials, this problem is potentially exacerbated
  by \llms trained on biased datasets.
However, \llms also offer an opportunity to address this issue. In
  this paper, we explore the use of \llms for automatically
  identifying gender-stereotypical elements in \Scratch tutorials,
  thus offering feedback on how to improve teaching content.
We develop a framework for assessing gender bias considering
  characters, content, instructions, and programming
  concepts. Analogous to how code analysis tools provide feedback on
  code in terms of code smells, we operationalise this framework using
  an automated tool chain that identifies gender \emph{stereotype
    smells}.
Evaluation on 73 popular \Scratch tutorials from leading educational
  platforms demonstrates that stereotype smells are common in
  practice. \llms are not effective at detecting them, but our
  gender bias evaluation framework can guide \llms in generating
  tutorials with fewer stereotype smells.
\end{abstract}



\maketitle

\section{Introduction}

\begin{figure}[t!]
\subfloat[Stereotypical girls
  project.\label{fig:example:girl}]{\includegraphics[width=0.42\columnwidth]{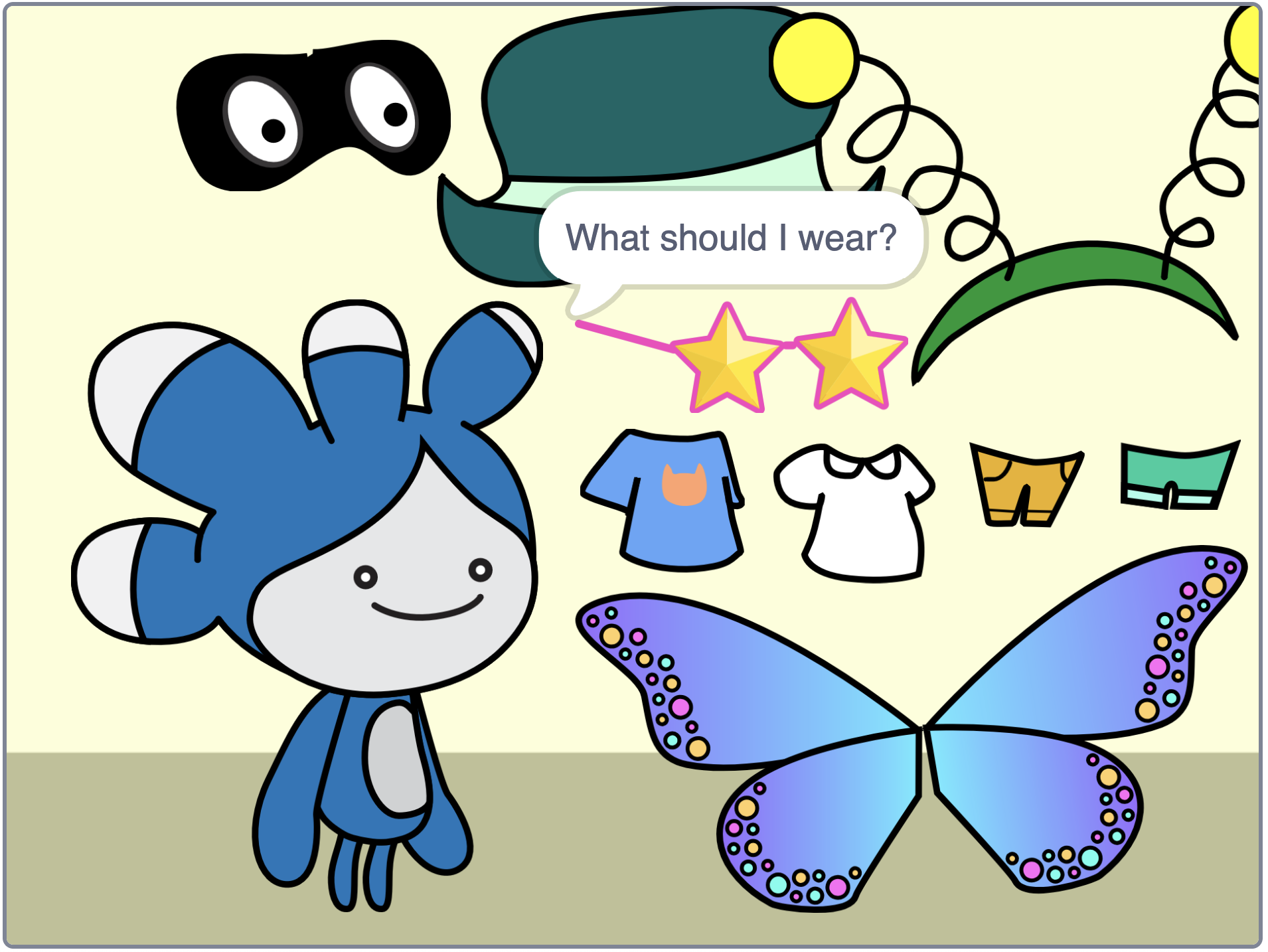}}
\hspace*{3ex}
\subfloat[Stereotypical boys
  project.\label{fig:example:boy}]{\includegraphics[width=0.42\columnwidth]{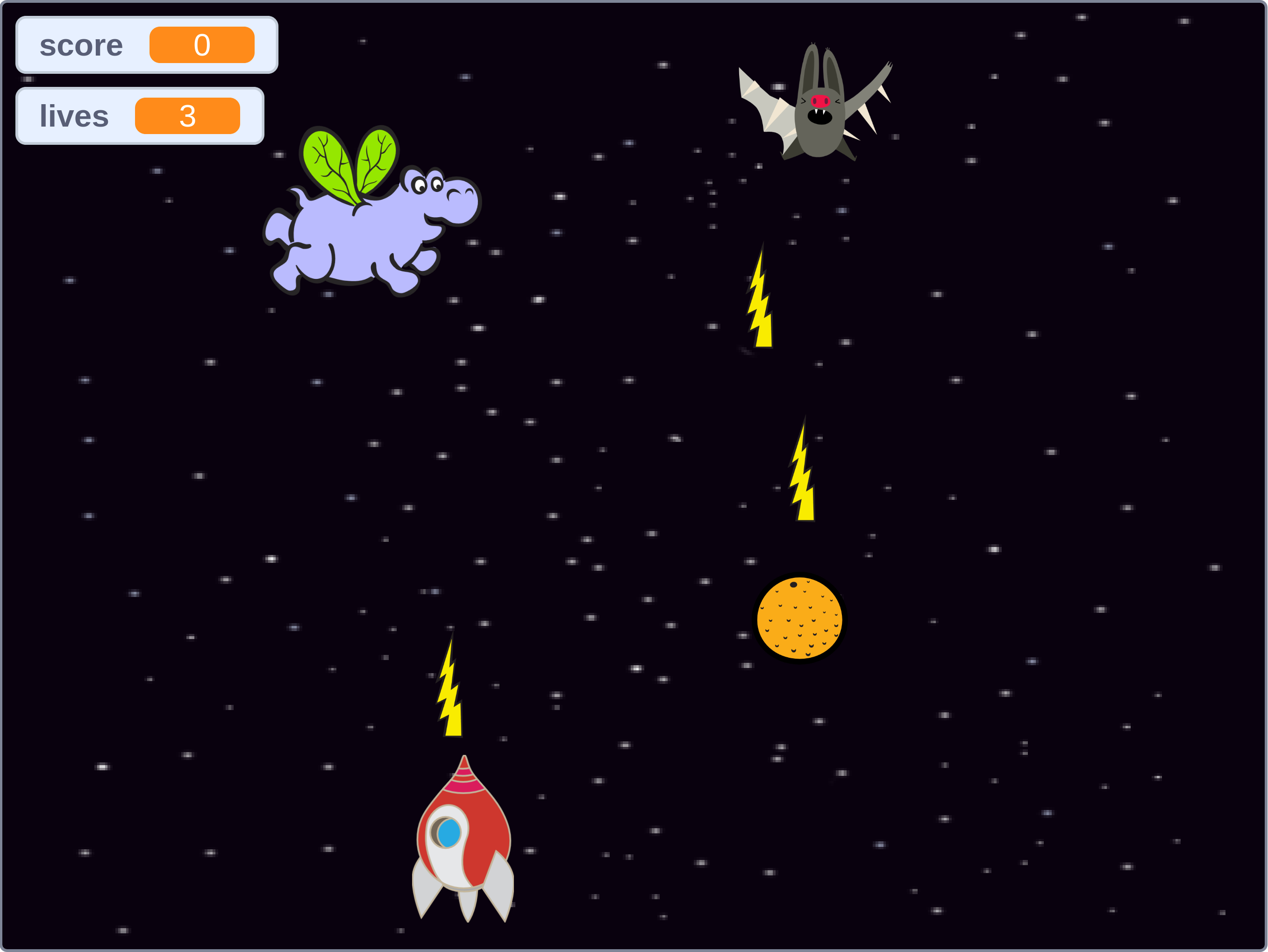}}
\caption{\Scratch programming tutorials with gender
bias.}\label{fig:example}
\end{figure}

Computing education faces a persistent challenge in creating inclusive
learning environments that attract and retain diverse young
learners~\cite{stattkus2025}.  Gender stereotypes in beginner
programming tutorials are often subtle and go unnoticed, yet they can
hinder learners' interest, learning outcomes, and
participation~\cite{schmidthaler2023a,greenwood2023,shenouda2024,spieler2020}.
These stereotypes manifest in various forms: from character
representations and narrative themes to colour choices and activity
types, all of which can unconsciously signal to students whether
computing is \emph{for them}~\cite{kafai2008}. For example,
\cref{fig:example} shows two \Scratch~\cite{resnick2009} tutorials
from the \emph{games} category:
\enquote{Dress Up Tera}
(\cref{fig:example:girl}), and \enquote{Clone Wars}
(\cref{fig:example:boy}).
The former reinforces gender stereotypes~\cite{bigler1992} by
targeting girls with traditionally feminine interests around
appearance~\cite{mahalik2005,shyian2021} while employing minimal
programming concepts, whereas the latter targets boys’ focuses on a
space shooter narrative with higher-level programming
concepts~\cite{spieler2020,kafai2008}.

While existing research has documented how such gendered patterns
manifest even in student-created \Scratch
projects~\cite{grassl2023,grassl2021,hsu2014}, there is a lack of
support for educators in identifying stereotypes in teaching
content~\cite{dewit2024}. This concern is exacerbated by the recent trend of using
Large Language Models (\llms) to generate educational
content~\cite{baidoo-anu2023, ali2024}, including programming
tutorials. Given that \llms may be trained on biased datasets they may
perpetuate or even amplify existing gender stereotypes in their
outputs~\cite{kotek2023, lucy2021}. Despite existing frameworks for
evaluating gender bias in educational materials~\cite{bhargava2002,
  heemskerk2005} and even tools that analyse textual gender bias in
educational resources~\cite{goritz2022}, there are currently no
approaches or tools that teachers could use to identify gender
stereotypes in teaching materials on \Scratch programming they use or
generate.

Besides their danger of projecting the gender biases of their training
data into teaching materials they generate, \llms, however, also offer
an opportunity to address this issue. In this paper, we explore the
use of \llms to automatically identify gender-stereotypical elements
in \Scratch tutorials. We consolidate existing research on gender
biases to create a framework for assessing programming tutorials, and
operationalise this in an automated tool chain that uses \llms to
assess \Scratch programming tutorials. Given a task description and
model solution, our approach provides educators with specific hints on
manifestations of gender biases in their teaching materials. Analogous
to how code analysis tools report \emph{code smells} as a means to
highlight issues of quality in the code, our approach reports gender
\emph{stereotype smells} that represent specific, actionable
manifestations of gender bias providing guidance for improving
programming tutorials. In addition, by referencing the framework in
the context provided to \llms, we can guide the creation of new
teaching materials to minimise the occurrence of stereotype smells.

To evaluate the proposed approach, we create a dataset of 73 
\Scratch programming tutorials from leading educational platforms as
well as 16 programming tutorials generated with an \llm. Using this
dataset, we investigate the prevalence of stereotype smells, and the
ability of \llms to identify them. We also evaluate whether our gender
bias evaluation framework can guide \llms in generating programming
tutorials that have fewer stereotype smells.

We find that one fifth of the 73 \Scratch tutorials contain stereotype
smells. In contrast to human evaluators, the \llm is not effectively
able to identify stereotype smells. When instructed explicitly, \llms
can produce \Scratch tutorials that contain fewer, but more nuanced,
stereotypes.  Gender-inclusive prompting commonly resulted in
reinforcing gender binarism, highlighting need for further
investigations of gender stereotypes in programming education.

 \section{Background and Related Work}

\subsection{Gender Stereotypes and Media}

\subsubsection{Gender and Stereotypes}
Gender is understood as a socio-cultural construct, in contrast
to the biological sex~\cite{rudman2021}. Gender inclusivity includes practices that support participation and representation of all genders.
Stereotypes are defined as beliefs about a certain group of
individuals' characteristics, attributes, and
behaviour~\cite{dovidio2010}.  These stereotypes can create psychological barriers that threaten
students' social identity and create feelings of incongruence between
student identity and educational
success~\cite{easterbrook2021,kizilcec2020}. These barriers can leave
members of underrepresented groups feel alienated and discouraged,
affecting their persistence and performance~\cite{stattkus2025,
  easterbrook2021}.
Gender stereotyping and consciousness already begins in early childhood~\cite{mckown2003}, with societal
norms and expectations shaping gender identities. Girls are more
likely to grow up with dolls and horses, but boys with superheroes and
ball sports~\cite{blakemore2005,hines2016}.

\subsubsection{Media and Video Games}
Gender stereotypes are often incorporated into self-regulation norms
through media products as a source of information on social roles,
which fosters gender-typed media use~\cite{bussey1999}. This research is
particularly relevant to programming environments like \Scratch, which are used to create various types of projects including games and
narratives~\cite{grassl2022e}.

Media representations often perpetuate stereotypes, with females depicted in subservient and passive roles~\cite{purcell1990}, only being interested in clothes and cosmetics~\cite{biraimah1993,chapman1997}, while boys are the active and being a leader or hero~\cite{purcell1990}. Despite shifts away from the damsel in distress narrative, much media still employs these stereotypes, especially for young audiences~\cite{elmously2024}. 
This young audience shows gender-dependent preferences in content: e.g., in movies and written stories, boys enjoy adventure, scary and violent scenes and tales more than girls, while girls prefer love stories and romantic tales~\cite{oliver2001,collins-standley1996,knobloch-westerwick2012,meier2025}.
Children also show preferences in their behaviour when interacting with multimedia stories, with boys paying more attention to navigational support, while girls pay attention to the colour and appearance~\cite{passig2000}. 
In video gaming, boys prioritise winning while girls see this as one of many options,
as they are more interested in taking their time to explore and 
discover~\cite{heeter2008}.  When given the choice in textbooks and animations, boys choose blue
and girls pink~\cite{karniol2011}, and girls prefer animal characters
independent of gender attributes, while boys prefer masculine
animals~\cite{karniol2000}.

\subsection{Gender Stereotypes in Computing}

\subsubsection{Stereotypes in Educational Software} 
Similar to other media products, educational software content typically
reflects male interest and is game-oriented with a focus on adventure,
sports, aggression, and
violence~\cite{calvert1999,biraimah1993,chapman1997,mangione1995}. Based
on constructivism, this will link children to the already intrinsic
stereotypes they have and thus reinforce existing
bias~\cite{calvert1999}. Girls, however, prefer girls as the central
figures of software~\cite{joiner1998,fiore1999}.
Thus, the interest of women in computing may be increased by avoiding
stereotypical patterns in courses~\cite{clarke2024}.
Gender-inclusive design therefore aims to avoid indicating
particular genders through stereotypical associations
for example by producing gender-neutral content. However,
designing gender-neutral educational games is
challenging~\cite{kafai2008,cassell2000}.
Various checklists have been developed to evaluate educational
software for gender bias~\cite{heemskerk2005,bhargava2002}, e.g., for
designing inclusive Information and Communication Technology~(ICT) materials~\cite{heemskerk2005} and to evaluate
educational software to promote girls’ interest~\cite{bhargava2002}. Automated approaches have adapted and
advanced those criteria~\cite{goritz2022}.

\subsubsection{Gender Bias in LLMs}

\llms also show bias regarding gender, social norms, and profession
expectations~\cite{kotek2023,zowghi2024}, reinforcing traditional
gender stereotypes with image generation models associating females
with family and appearances while men are portrayed with power and
professions~\cite{lucy2021,sun2024,lucy2021}, e.g., software
developers predominantly being male~\cite{bano2025}.  The absence of
representation from marginalised groups risks reinforcing stereotypes
and inequalities~\cite{bender2021}.
This is especially true for educational material as it is about the
norms that lie beyond the \emph{`surface narrative, the
  backstory'}~\cite{kafai2008}.

\subsubsection{Gender Stereotypes in Scratch}
\Scratch is a block-based programming language designed to teach programming to children with creativity as a core concept through sprites and stages using different blocks that can be combined and applied to different programming concepts~\cite{roque2016,resnick2009}. 
Prior studies on gendered patterns in \Scratch identified issues
impacting self-efficacy, interest,
creativity~\cite{grassl2023,grassl2021}. Subtle stereotypes in the
projects children create can influence the use of blocks and thus the
applied programming concepts, resulting in lower quality and
complexity of girls' code, broadening the prior knowledge gap in
computing~\cite{grassl2023,spieler2020,hsu2014}.

\section{Research Design}

\Scratch tutorials are widely used in K-12 programming education, yet
systematic evaluation of gender bias in these materials remains
limited. Thus, our first research question is as follows:

\noindent\textbf{RQ1:} Do common \Scratch tutorials contain
stereotypes?

Educators need reliable tools to identify potentially biased content
that may discourage girls from engaging with programming.  We
therefore examine whether \llms can serve as effective tools for bias
detection, and our second research question is as follows:

\noindent\textbf{RQ2:} Can \llms detect gender bias in
\emph{existing} \Scratch tutorials?

While \emph{RQ1} and \emph{RQ2} examine bias detection in existing
content, educators might increasingly use \llms also to
\emph{generate} programming tutorials. However, the gender bias
present in \llm-generated content and educators' ability to identify
such bias in materials remains unexplored. Thus, our third research
question is as follows:

\noindent\textbf{RQ3:} How do gender-inclusive prompting strategies
affect gender bias in \llm-\emph{generated} \Scratch tutorials as
assessed by educators?

\subsection{Data Generation}

Our data collection follows a two-step approach: first, we analyse
existing official \Scratch tutorials from two established educational
platforms; second, we generate new tutorials using different prompting
strategies to create a diverse dataset for testing \llm bias.

\subsubsection{Existing Educational Tutorials}\label{subsubsec:tutorials}
We collected 75 \Scratch tutorials from two established educational platforms to represent state of the art educational programming content for beginners. 
These platforms were selected based on their prominence in educational
programming instruction and their widespread adoption in schools
globally. The \Scratch tutorials vary in project type, providing a
representative sample of current educational content.

\textbf{Scratch.org} represents the official educational content from
the creators of \Scratch, ensuring authenticity and pedagogical
alignment with the platform's educational objectives. 
We collected all 40 starter projects from the official
website,\footnote{\url{https://scratch.mit.edu/starter-projects},
accessed 2025-07-11} consisting of eight categories with five projects
each: \emph{animation, games, interactive art, music, stories, math
and science, extensions, community and kindness}.

\textbf{Raspberry Pi Foundation} provides comprehensive,
curriculum-aligned programming resources that are extensively used in
formal and informal educational settings, particularly in K-12.
We collected all 35 projects from the official
website.\footnote{\url{https://projects.raspberrypi.org/en/collections/scratch},
accessed 2025-07-11}

We collected the executable \Scratch projects (.sb3 format) together with
metadata including
instructions and project category.
Since two project links did not work, we excluded them from our
analysis, resulting in a dataset of 73 tutorials. We provide all
materials in our replication package~\cite{KoliReplicationPackage}.

\subsubsection{LLM-Generated
Projects}\label{sec:data-gen:phase2}

To test whether \llms introduce gender bias, we generated 16 \Scratch
project descriptions using the prompt shown in
\cref{fig:prompt-project-gen}. We generated two projects for each
category (\emph{`project-topic'}) based on the \Scratch official
website categories~(cf.~\cref{subsubsec:tutorials}).  For each
category, we tasked the \llm to describe a project without mentioning
that the tutorial should be gender-inclusive to establish a baseline
of what \llms produce.
We then also queried the \llm with the same prompts again, but this
time explicitly requested the project to be gender-inclusive using the
gender bias evaluation framework described in \cref{sec:framework}.

We used \gpt{}\footnote{\url{https://openai.com/index/gpt-4-1/},
accessed 2025-07-14} version gpt-4.1-2025-04-14 via the official
\openai{} \api for all experiments. The model is multimodal and
supports both text and images in the input. Although the output of
\llms can vary due to its non-deterministic behaviour, we observed a
sufficient degree of consistency in concepts included across generated
projects. To ensure scalability of the human rating effort, we
therefore include only a single project per prompt.

\begin{figure}[t]
  \fbox{\parbox{0.96\linewidth}{\footnotesize I am teaching programming to children. We use the Scratch programming language.
I need help to create a Scratch project the children can implement to learn basic programming concepts.

\textbf{if} \emph{gender-inclusive} \textbf{then}
The project should be gender-inclusive.
As a guideline, researchers have identified the following criteria to check whether a project is gender-inclusive:
\emph{list of framework statements}
\textbf{endif}

If it makes sense, the project should contain basic programming concepts like conditions and other control structures.
Keep in mind that the children are beginners. Do not include too many programming concepts at once.

The general topic for the project should be: `\emph{project-topic}'

Describe a Scratch project that fulfils the requirements above.
Focus on the project design, features, and programming concepts in your description.
I also want to demonstrate the children an example of this project at the beginning of the class to get them motivated.
Describe which sprite and background images I should choose for this example demonstration.
You should not output any code. The project and example image descriptions are enough.
    }
  }
  \caption{Prompt for RQ3. Placeholders in \emph{italics}. To request inclusive projects, we include the statements from \cref{tab:checklist-stereotypes}.}\label{fig:prompt-project-gen}
\end{figure}

\subsection{Scratch Gender Bias Evaluation
Framework}\label{sec:framework}

\begin{table}[t]
\centering
\caption{Criteria of the Gender Bias Evaluation Framework.}\label{tab:checklist-stereotypes}
\resizebox{\linewidth}{!}{\footnotesize \setlength{\tabcolsep}{2pt}
\begin{tabular}{lp{6.9cm}l}
\toprule
 & Statement & Source \\
\midrule
\multicolumn{3}{l}{Characters} \\
CH01 & Female and male characters are equally represented. & \cite{bhargava2002} \\
CH02& Active and passive behaviors are equally distributed between female and male characters. & \cite{bhargava2002}\\ CH03& Female characters are presented in problem solving \& leadership roles. & \cite{bhargava2002}\\
CH04& The characters are not stereotyped by occupational roles. & \cite{bhargava2002}\\ CH05& The types of emotional statements attributed to females and males are not stereotypic. & \cite{bhargava2002}\\ 

\midrule
\multicolumn{3}{l}{Content} \\

CO01& The project is free of sexist language. & \cite{bhargava2002}\\
CO02& The extent and frequency of aggressive and/or destructive behaviors is limited or non-existent. & \cite{bhargava2002}\\
CO03& Content and overall style of the project appeal to girls \& boys. & \cite{bhargava2002}\\
CO04& The preferences of different groups are taken into account in the visual interface (e.g. bright vs dark colours). & \cite{heemskerk2005} \\
CO05& Non-human objects are not assumed to be male. & \cite{bhargava2002}\\
CO06& The number of elements of competitiveness vs. cooperation is balanced. &\cite{bhargava2002} \\
CO07 &Themes and scenarios reflect diverse interests beyond stereotypical gender preferences.& [new]\\

\midrule
\multicolumn{3}{l}{Instructions} \\

IN01 & The project addresses different kinds of skills (e.g. writing, drawing). &\cite{heemskerk2005} \\ 
IN02 & The project is made in such a flexible manner that students can alter parts to their preference (e.g. own sprites). &\cite{heemskerk2005} \\ \midrule

\multicolumn{3}{l}{Programming Concepts} \\

PR01 & Basic programming concepts (i.e., loops, conditionals) are used. & [new]\\ PR02& The project complexity is appropriate for beginners.& [new]\\
PR03&The project presents programming as having diverse applications beyond games and competition.& [new]\\
PR04 &Creative and artistic programming features are balanced with logical or computational elements. & [new]\\
\midrule
\multicolumn{3}{l}{Verdict} \\
VE01 & Is the program targeted towards boys, girls, or is it gender-inclusive? & [new]\\
\bottomrule
\end{tabular}
}
\end{table}

Providing actionable feedback on gender bias requires specific
criteria that teaching material can be checked against. As there
is no such framework specifically for \Scratch programming tutorials
and their particular opportunities for gender bias, we consolidated
existing work on criteria for educational software on gender bias
assessment and adapted it to the specific context of \Scratch.

We draw on two foundational frameworks: The first~\cite{bhargava2002}
identified three key areas in educational textbooks and subsequently
adapted them to software: (1)~characters, (2)~content, and (3)~rewards.
Since reward systems may only apply to specific games we excluded this
aspect.
To identify further relevant items for gender bias and to verify items
in the first framework, a second framework~\cite{heemskerk2005} was
utilised which focuses on (1)~the content of ICT tools, (2)~their
visual and audio interface, and (3)~their instructional structure.
Several content-related aspects, such as the presence, representation,
and contextualisation of different groups, overlap with the first
framework, and we adopted that phrasing due to its conciseness. Items
containing explicit instructions or feedback mechanisms are targeted
towards application in a classroom setting, but not applicable in our
context, so we did not include them. 
Bhargava~\cite{bhargava2002} used a scale of \emph{often},
\emph{sometimes}, \emph{never}, with an additional option \emph{not
  applicable}. The second framework~\cite{heemskerk2005} does not
propose a specific scale. We adapted our scale towards a measure of
intensity rather than frequency, represented by a 5-point Likert scale
plus \emph{not applicable}.
I.e., \emph{often} is a strong presence of the respective criterion
(strongly agree), \emph{sometimes} is a moderate presence (neither
agree nor disagree), and \emph{never} is an absence (strongly
disagree).

\Cref{tab:checklist-stereotypes} summarises our adapted framework. It
captures both explicit gender representation (e.g., character
distribution and roles) and implicit biases (e.g., assumptions about
diverse skill sets, emotional attribution).
In cooperation with an interdisciplinary team of computer scientist,
(computing) educators, and experts in gender studies, we introduced
\emph{CO07} capturing the themes and scenarios since \Scratch is
designed for creative storytelling~\cite{roque2016} as well as four
novel criteria (\emph{PR01--PR04}) to assess whether the used
programming concepts might implicitly favour certain learner
groups~\cite{grassl2021,hsu2014,spieler2020}.

\paragraph{Characters (CH01--CH05)}
Character representation directly impacts whether students see
programming as \emph{for them}. Since sprites are the primary characters that students program and
interact with, gender balance in sprite selection (\emph{CH01}), equal
assignment of active behaviours (\emph{CH02}) and leadership roles
(\emph{CH03}) counter the stereotype that protagonists are
predominantly male.  Occupational stereotyping (\emph{CH04}) and
emotional expressions (\emph{CH05}) are particularly relevant as in
\Scratch often involves role-playing scenarios.

\paragraph{Content (CO01--CO08)}
\Scratch tutorials often involve concepts like storytelling, game
creation, interactive art, and stylistic elements.
Gender-neutral (\emph{CO01}) character names, dialogue, and project descriptions
prevent alienation of any group. Violence levels (\emph{CO02}) matter
because many \Scratch projects involve character interactions and
conflict resolution. Visual preferences (\emph{CO04}) affect appeal,
as tutorials dominated by dark, aggressive or overly \emph{cute} or
\emph{shiny} designs may discourage certain students. The balance
between competitive game mechanics and collaborative storytelling
(\emph{CO06}) shapes students' perception of programming. Avoiding
male-default assumptions for non-human objects (\emph{CO05}) and
presenting diverse themes (\emph{CO07}) helps frame programming as a
creative and practical, rather than just technical, tool for
problem-solving.

\paragraph{Instructions (IN01, IN02)}
This category addresses how tutorials accommodate different learning
backgrounds and preferences, since \Scratch tutorials should guide
students through programming concepts without assuming prior
knowledge.
Addressing multiple skill types (\emph{IN01}) acknowledges that
\Scratch programming does not just involve technical programming, but
also creative designing, (logical) sequencing, and creative
storytelling, as well as adapting it to the students preferences
(\emph{IN02}), e.g., by adding their own sprites.

\paragraph{Programming Concepts (PR01--PR04)}
This category addresses how technical content is presented and whether
it reinforces or challenges gendered skill gaps.
The inclusion of basic programming concepts like loops and
conditionals~(\emph{PR01}) is relevant since avoiding these concepts
may unconsciously limit students' development. Gendered projects
lacking such concepts~(e.g., the basic girls-centred animation
compared to the boys-centred game shown in \cref{fig:example}) are
especially harmful since they may result in gendered skill gaps.
Complexity levels require careful consideration since girls often have
less prior programming experience, making tutorials that assume
advanced knowledge systematically exclusive, whilst
beginner-appropriate complexity ensures equitable access (\emph{PR02}).
Presenting programming only through games and competition reinforces
narrow perceptions about who computing is for, whereas using diverse
applications (\emph{PR03}) and balancing between creative (art, music,
animation) and logical (conditionals, loops) blocks (\emph{PR04})
demonstrates programming addresses any interests.

\paragraph{Verdict (VE01)}
One criticism of the original framework~\cite{bhargava2002} is that it
is not clear in which direction the gender bias lies, as only the
authors' rather the teachers' statements whether projects are targeted
towards girls or boys are used. Therefore, we introduce a question of
a verdict (\emph{VE01}) to capture the judgement of the raters.

\subsection{Evaluation Process}

For all evaluation steps that involve an \llm, we again use the same
\gpt variant as for the data generation
step~(cf.~\cref{sec:data-gen:phase2}).

\subsubsection{Evaluation Criteria}

To evaluate the human and \llm responses, we computed the following
metrics based on our framework:

\noindent\textbf{Average criterion scores \(\varnothing\):}
For each criterion of the framework, we aggregated the rater responses
given on a 5-point Likert scale and divided by the number of raters to
obtain a mean score per criterion~\cite{bhargava2002}.
The criteria statements are formulated such that a lower criterion
score indicates a less stereotypical program.

\noindent\textbf{Overall framework scores \(\sigma\):}
The overall framework score \(\sigma\) for a project is determined by
the mean of all individual 5-point Likert scale framework score
ratings, i.e., the mean over \(18 \cdot n\) scores in the 18 framework
criteria from \(n\) raters~\cite{bhargava2002}.

\noindent\textbf{Overall verdict:}
For each project, both human and \llm raters gave an overall verdict
whether the project is targeted at boys, girls, or if it is
gender-inclusive~(\emph{VE01}).
In the evaluation, we focus mainly on projects being identified as
gender-specific by at least one rater since this is a clear indicator
of potential gender bias in the project.
By checking that projects identified according to this verdict also
have a high overall framework score, we can ensure our framework
reflects the relevant criteria raters use to form their decision.
Using the average criterion scores, we can then further identify which
criteria were important for the overall verdict.

\subsubsection{RQ1}

Each existing \Scratch tutorial was independently evaluated by six
computing education researchers selected through convenience sampling
due to the need for participants with relevant expertise in \Scratch,
computing education, and basic awareness of stereotypes as well as
availability.  They all have experience in (computing) education
and/or research in this field, and were informed about the purpose of
the study. The raters were presented with a list of the 73 \Scratch
projects (cf. \cref{subsubsec:tutorials}), and tasked to look at each
\Scratch tutorial on its website
and rate it according to the framework
categories~(\cref{tab:checklist-stereotypes}).
To answer RQ1, we analysed both scores and the overall verdicts of
human evaluators, focusing on projects identified to be gendered.

\subsubsection{RQ2}

\begin{figure}[t]
  \fbox{\parbox{0.96\linewidth}{\footnotesize I am teaching programming to children. We use the Scratch programming language.
I have found a Scratch project I would like to use as a starter project. The children then extend it in class with their own ideas.

Here is the Scratch program in the ScratchBlocks format as you know it from the Scratch community forums:
\emph{scratchblocks-code}\\
It was described by the creator as: \emph{description}\\
I also give you the images of the sprites of the project and a screenshot of the whole stage as it appears at the start of the program.

\textbf{if} \emph{with-framework} \textbf{then}
Researchers have identified the following checks to identify if a program is gender-inclusive:\hspace{3ex}
Identifier\hspace{2ex}Description\\
\hspace*{34.7ex}CH01\hspace{5.5ex}…

Keep in mind that not only humans but also non-humans, animals, or objects can be characters and have gender-specific features.
Answer the following question:\\
Is the program targeted towards boys, girls, or is it gender-inclusive?

Use the checks from the table above to form your answer.
Answer each check on a five-point Likert scale ranging from 1=strongly agree to 5=strongly disagree, 3 representing neither agree nor disagree, or alternatively 0=not applicable.
Output your answers as a list of the following format:
\texttt{<Identifier>: <score>}.
Use the \texttt{<Identifier>} from the first column of the table above and \texttt{<score>} as number between 0 and 5 on the Likert scale.
Finally, give your conclusion as \enquote{@boy@}, \enquote{@girl@}, or \enquote{@inclusive@}.
\textbf{else}

Is the program targeted towards boys, girls, or is it gender-inclusive?
Keep in mind that not only humans but also non-humans, animals, or objects can be characters and have gender-specific features.
Explain your answer.
Give your final answer as \enquote{@boy@}, \enquote{@girl@}, or \enquote{@inclusive@}.
\textbf{endif}
    }
  }
  \caption{Prompt for RQ2. Placeholders in \emph{italics}.}\label{fig:prompt-has-stereotypes}
\end{figure}
 
The \Scratch sb3 files contain the project code in \json format and
necessary resources like sprite images and sounds. We do not use this
\json as part of our prompts to the \llm directly but instead use
\litterbox~\cite{Fraser2021} to convert the program into the
\scratchblocks
format\footnote{\url{https://en.scratch-wiki.info/wiki/Block_Plugin/Syntax},
accessed 2025-07-24} which has likely been seen by the \llm during
training, since it is used on the \Scratch community forums.
Additionally, we extract the costume images of the sprites and a
screenshot of the program stage when starting the program and send
them in addition to our
prompt~(cf.~\cref{fig:prompt-has-stereotypes}). Including the sprite
images in our request ensures that program features not appearing in
the code, like for example the colour of objects, are taken into
account.

We prompt the \llm with two prompt variants. The first one does not
include the framework~(\emph{else} case in
\cref{fig:prompt-has-stereotypes}). This elicits whether the \llm has
learned to reason about the inclusiveness of a program without having
concrete guidelines it can check. To encourage a chain-of-thought
process~\cite{Wei2022} even without the framework, we request it to
explain the answer. The second prompt variant includes the framework and asks the \llm to
rate the project on a 5-point Likert scale for each criterion.
For both prompt variants, the \llm responded consistently in the
requested format.
To account for the inherent randomness in the \llm{}’s output, we
repeat each prompt five times and treat the responses as if given by
five different raters.

To answer RQ2, we extract the overall verdicts from the responses in
both prompt variants to identify projects classified by the \llm as
gendered.
By also extracting the framework scores from the second prompt
variant, we can compare \llm-generated scores~(\(\sigma_m\), \mmean{})
with human evaluation scores~(\(\sigma_s\), \hmean{}) from RQ1 both
for the overall framework scores and for the individual criterion
scores.

\subsubsection{RQ3}

For the \llm-generated \Scratch tutorials, six educators evaluated the
projects according to the same framework and overall verdict as used
in the evaluation of RQ1 and RQ2. We additionally asked them whether
they would use the program as-is, with small or with larger
modifications, or not at all in their classroom in order to get
insights of the practical use of the generated tutorial. We set up an
online study displaying the description of each generated tutorial and
the items of the framework. We informed the educators about the aim of
the study. However, they were not aware of which version of the 16
tutorials had been prompted with/without gender-inclusiveness. To
avoid further biasing, we constructed three questionnaires with
different orders of the projects.

To address RQ3, we analysed scores and verdicts from human evaluation
of the 16 generated projects, comparing stereotype smells between
explicitly gender-inclusive projects and their standard counterparts
without gender-inclusive prompting.

\subsection{Data Analysis}

To ensure validity, we implemented multiple validation mechanisms:
comparison of \llm outputs to human expert annotations, mapping
identified stereotypes to literature-derived categories from our
framework, and assessment of inter-rater reliability~(IRR, computed
using Fleiss' \(\kappa\)~\cite{Fleiss1971}) amongst multiple coders.
Human evaluations involved six human raters, whilst for the \llm, we
used the five \llm responses per project to calculate reliability for
overall scores.

\subsubsection{RQ1}

Out of the six raters, four identified as female and two as male, all
from Germany. The raters ranged in age from 25 to 32 years
($M = 27.7, SD = 3.1$) and included four researchers and two students
across the fields of software engineering, computer science, and
education. Three raters have taught at university level, at secondary
level, and at primary level each, with some raters teaching across
multiple levels. Computing education experience varied considerably.
The raters demonstrated high familiarity with \Scratch, with ratings
on a 5-point scale (1 not familiar at all, 5 expert level) ranging
from 2 to 5 ($M = 4.2, SD = 1.2$) as well as a moderate level of
confidence in rating stereotypes in \Scratch projects, with ratings
ranging from 2 to 4 ($M = 3.2, SD = 0.8$). The perceived applicability
of the stereotypes evaluation framework for the existing 73 projects
was moderate with ratings ranging from 2 to 4 ($M = 2.6, SD = 0.9$).
The raters reached moderate agreement~\cite{landis1977measurement} on
the same overall verdict for the
projects~(\fkappa{0.5780821917808219}), but fair agreement on the
criteria of the framework~(\fkappa{0.3137899543378995}).
This variation may be explained by differences in awareness of
stereotypes: two female raters demonstrated greater sensitivity and
identified more implicit stereotypes, whereas the remaining four
raters tended to focus on more explicit features such as gendered
colour schemes or stereotypical narratives.

\subsubsection{RQ2}

The \llm achieved almost perfect agreement for both the basic prompt
variant~(\fkappa{0.967123287671233}) and the one including the
framework~(\fkappa{0.9917808219178079}). It showed substantial
agreement in scoring the projects according to the framework’s
criteria~(\fkappa{0.7060273972602741}).

\subsubsection{RQ3}

Four female and two male raters, aged 25 to 42 years ($M = 32.7, SD =
6.6$), participated. Three raters were from Germany, two from
UK, and one from Canada. Two raters have taught at
university and primary level each, and six at secondary level (with
some raters teaching across multiple levels). Two female
researchers/students from the RQ1 evaluation also participated in the
RQ3 evaluation to maintain consistency and enable comparison across
research questions.
Computing education experience was substantial, ranging from 3 to 20
years ($M = 10.3, SD = 6.9$). All raters demonstrated good to high
familiarity with \Scratch, with ratings on a 5-point scale ranging
from 3 to 5 ($M = 4.0, SD = 0.9$). Confidence in rating stereotypes in
\Scratch projects was consistently moderate to high, with all ratings
at 3 or 4 ($M = 3.7, SD = 0.5$). The perceived applicability of the
stereotypes rating framework was notably higher than in RQ1, with
ratings ranging from 3 to 5 ($M = 4.2, SD = 0.8$).
The raters reached substantial agreement on the usability of the
project descriptions~(\fkappa{0.6888888888888888}), almost perfect
agreement on the final inclusiveness
verdict~(\fkappa{0.8812499999999998}), and moderate agreement on the
individual framework categories~(\fkappa{0.5672222222222222}). The two
female raters with prior experience in stereotype analysis
demonstrated higher sensitivity to nuanced forms of bias, while the
other raters tended to largely identify explicit stereotypes.

\subsection{Threats to Validity}

\paragraph{Internal}
Potential systematic biases in \llm training data may affect the
accuracy, leading to consistent over- or under-detection of certain
stereotype categories. We employed a state-of-the-art \llm and
repeated the prompts to account for inherent randomness.
Our prompting strategies may not capture all possible variations,
potentially missing subtle or context-dependent stereotypes. However,
we iteratively refined prompts through pilot testing and employed
multiple prompt variations to ensure best possible coverage.

\paragraph{Construct}
The operationalisation of gender stereotypes within our framework may
not fully capture the multidimensional nature of stereotypical
representations in programming contexts. However, our framework is
grounded in established literature on gender stereotypes in computing
education and validated through expert review and pilot testing with
educational materials. Given the partially moderate IRR in our study,
future work should explore refinements to the framework criteria,
potentially developing \Scratch-specific categories through focus
groups with educators.
Variations in how human raters and \llms interpret stereotype
categories may lead to inconsistent measurements, and bias may result
from the convenience sampling since stereotypes might differ between
cultural and social backgrounds. We report IRR measures to ensure
measurement consistency and subjective in nature. Future studies
should aim to recruit raters with more varied cultural and
disciplinary backgrounds and with different levels of expertise in
stereotype analysis, in order to increase robustness and sensitivity.
All data is
available in our replication package~\cite{KoliReplicationPackage} for
further validation.

\paragraph{External}
The sampled \Scratch tutorials may not represent the full diversity of
programming materials as well as different representation modalities,
limiting generalisability. However, they represent current best
practice in programming education.
Our findings lack validation through actual classroom implementation,
limiting understanding of real-world impact and student responses to
identified stereotypes. We envision future research on classroom
studies and student feedback to validate the practical significance of
identified stereotype smells in real-world learning environments.

\subsection{Positionality Statement}

We recognise that positionality is central when interpreting bias and
stereotypes, as such judgements are never neutral. All authors have
years of experience in computing education and in teaching \Scratch,
which gives us awareness of explicit stereotypes such as gendered
narratives, colour schemes, or role assignments. Our combined
expertise spans software engineering, machine learning and \llms, and
the humanities with gender studies. These perspectives let us examine
stereotypes in \Scratch from technical, educational, and
socio-cultural angles.
The team identifies as one woman and two men, with a shared European
background that shapes how we interpret gender norms and stereotypes,
which may differ from other cultural contexts. While we did not
attempt to standardise these perspectives, we view the variation as
reflecting the diversity among computing educators and learners.

 \section{Results}

\subsection{RQ1: Gender Bias in Scratch Tutorials}\label{sec:rq1}

The majority of the 73 official \Scratch tutorials were rated as
gender-inclusive by human evaluators in their overall verdict.
However, 19.18\,\%~(14 projects) were evaluated as being stereotyped,
nine projects explicitly targeting girls and five projects targeting
boys.

\subsubsection{Character Representation}

The most pronounced stereotypes appear in character categories
(\emph{CH01--CH04}), with scores consistently ranging from 3.5 to 5.0.
This is mainly influenced by the central sprites of these projects and
their roles.
In particular, \emph{Dress Up Tera} (Scratch project ID 1105678528,
\score{2.9}, shown in \cref{fig:example:girl}) stands out as the
project with the highest overall stereotype score, with particularly
high character category ratings (\emph{CH01--CH04}, values ranging
from 4.0 to 4.5). The project centres on customising a
female-associated character named \emph{Tera} and selecting clothing
and accessories like sunglasses, focusing on the traditionally
feminine question of \emph{what to wear}. This reinforces
sociocultural associations between women and appearance, where social
capital derives primarily from beauty rather than
achievement~\cite{mahalik2005,shyian2021}.

Various manifestations of character bias towards girls can be found in
\emph{Memory} (ID 34874510, \score{2.8}), featuring a \emph{Ballerina}
dancing on a stage while her tutu changes throughout the project
(\emph{CH01} \mean{5.0}, \emph{CH02} \mean{4.6}, \emph{CH04}
\mean{4.3}), or \emph{Story Starter} (ID 1110565816, \score{2.8}), which features
three girl characters (\emph{Trisha, Tatiana, Taylor}) whose primary
activity is talking to each other (\emph{CH01} \mean{4.8}, \emph{CH02}
\mean{4.2}). This reinforces stereotypes about female communication
patterns without demonstrating any other interests, activities, or
capabilities~\cite{hines2016}.
In \emph{Broadcasting Spells} (ID 518413238, \score{2.7}) the player
employs a magic wand to shrink, grow, or transform a fairy and a
female football player into toads while retaining their costumes,
demonstrating passive roles (\emph{CH01} \mean{4.2}, \emph{CH02}
\mean{3.6}, \emph{CH03} \mean{3.5}).
In \emph{Random Acts of Kindness} (ID 1110573738, \score{2.5}) the
female character \emph{Dani} wonders \enquote{Hmm\textellipsis{} What
can I do to be kind?}, and receives suggestions from an \emph{idea
bulb} sprite including \enquote{I could help my family with a
household chore} or \enquote{I could clean up my workspace}
(\emph{CH01} \mean{4.2}, \emph{CH02} \mean{4.2}), showing an emphasis
on domestic and caring roles~\cite{hines2016,lucy2021}.
Finally, in \emph{Text to Speech} (ID 1106234816, \score{2.4}) the
sprite \emph{Abby} uses a \emph{squeaky} voice (explicitly female) to
converse with a second character, the \emph{Chick}, which responds in
a giant voice.

On the other hand, projects flagged as biased towards boys feature
characters that embody aggression and dominance, even when characters
are non-human.
\emph{Clone Wars} (ID 46018140, \score{2.8}) achieves the
second-highest stereotype score overall and the highest among
boy-targeted projects. Users control a spaceship with the aim of
shooting down a bat with blasts against a space background. Despite
featuring no human characters, evaluators perceived them as male
(\emph{CH01} \mean{3.5}).
\emph{Swarms Schools and Flocks} (ID 547542437, \score{2.7}) presents
a game where users control a bat with clones, losing clones when
touching dinosaurs or lions, with complete clone loss resulting in
death and game over. The project scores are high across multiple
character categories (\emph{CH01} \mean{3.8}, \emph{CH02} \mean{3.8},
\emph{CH03} \mean{5.0}, \emph{CH05} \mean{5.0}), emphasising survival
through combat and dominance over other creatures.
\emph{Ghostbusters} (ID 60787262, \score{2.6}) places users in a dark
forest to catch ghosts by clicking them to score points (\emph{CH01}
\mean{4.0}, \emph{CH02} \mean{3.7}), reinforcing associations between
masculinity and confronting supernatural threats.
\emph{DJ Scratch Cat} (ID 11640429, \score{2.4}) presents the \Scratch
cat as a hip-hop DJ playing music, positioned at turntables with large
speakers. The character embodies stereotypical masculine urban culture
aesthetics (\emph{CH02} \mean{3.7}, \emph{CH03} \mean{5.0},
\emph{CH04} \mean{3.8}, \emph{CH05} \mean{4.0}).
Finally, \emph{Interactive Parallax} (ID 1105131011, \score{2.3})
features a dinosaur world where clicking sprites changes them to
different dinosaurs (\emph{CH01} \mean{3.5}, \emph{CH03} \mean{3.7}).
The prehistoric theme and creature transformation mechanics appeal to
stereotypically masculine interests in power and control.

\subsubsection{Content and Themes}

Projects targeting girls frequently focus on traditionally feminine
interests, scoring high on content appropriateness (\emph{CO03}) and
stereotypical theme reinforcement (\emph{CO07}).
\emph{Dress Up Tera} may not appeal to both girls and boys
(\mbox{\emph{CO03} \mean{4.2}}) as it reinforces stereotypical gender
preferences learned from society (\emph{CO07} \mean{4.3}). Similarly,
\emph{Story Starter} scores high on both measures (\emph{CO03}
\mean{4.2}, \emph{CO07} \mean{3.8}) by simulating stereotypical
behaviour of girls talking amongst themselves.
\emph{Flower Generator} (ID 253355932, \score{2.52}) requires users to
generate flowers, which may not represent diverse interests since
flowers and nature are associated with femininity (\emph{CO07}
\mean{3.8}). 
\emph{Food Truck Animation} (ID 1105114421, \score{2.5}) features a
pink-coloured food truck, reinforcing gender colour associations
(\emph{CH01} \mean{4.0}, \emph{CH02} \mean{4.5}, \emph{CH03}
\mean{4.5}, \emph{CO03} \mean{4.0}).
\emph{Mandala} (ID 536953224, \score{2.3}) asks users \enquote{What
  feels most peaceful to you today? (1) abstract shapes (2) the earth
(3) butterflies (4) love}, emphasising emotion and aesthetics rather
than other considerations.

On the other hand, competition seems to dominate the design philosophy
of tutorials biased towards boys, with rather destructive mechanics
central to user engagement.
In particular, \emph{Clone Wars} exhibits the highest destructive
behaviour score (\emph{CO02} \mean{3.2}) amongst all tutorials and
demonstrates omnipresent competitive elements (\emph{CO06}
\mean{4.3}), where success depends entirely on defeating enemies.

\subsubsection{Instructions and Programming Concepts}

A concerning pattern emerges in programming complexity, with projects
targeting girls consistently showing simplified programming
structures.
\emph{Dress Up Tera} contains mostly sprites with only sequences,
meaning the project would work successfully without any loops or
conditionals~(\emph{PR04} \mean{3.3}).
\emph{Story Starter} similarly employs predominantly sequential
programming with only one repeat block (\emph{PR04} \mean{3.5});
\emph{Memory} and \emph{Broadcasting Spells} (\emph{PR01} \mean{3.8})
use only basic sequences except for one loop.
\emph{Text to Speech} uses only sequences without any complexity or
control blocks at all (\emph{PR01} \mean{4.0}).
\emph{Flower Generator} raises questions about whether the complexity
level is appropriate for beginners (\emph{PR02} \mean{4.2}),
suggesting either oversimplification or unclear instruction design.
\emph{Random Acts of Kindness} features sprites with mostly easy
sequences, except for having one block using lists (\emph{PR01}
\mean{3.5}).
This could reinforce often-reported initial lower programming
experience amongst girls in courses~\cite{spieler2020}.  However, the
pattern of reduced complexity in projects targeting girls may limit
their programming learning opportunities, potentially reinforcing
rather than addressing documented gender gaps in programming
self-efficacy and experience.

Projects targeting boys show limited adaptability to diverse student
preferences and interests. The rigid game mechanics in \emph{Clone
Wars}, \emph{Swarms Schools and Flocks}, and \emph{Ghostbusters} offer
little room for creative adaptation or alternative interaction styles.
These projects consistently score high on inflexibility measures
(\emph{IN02} values between \mean{4.0} and \mean{4.2}), e.g.,
\emph{Clone Wars }highlights the inflexible shooter game concept
reinforcing the stereotype that programming equals gaming (\emph{PR03}
\mean{4.2}).
\emph{Swarms Schools and Flocks} similarly emphasises survival
competition, with users constantly threatened by predators and
required to maintain populations through strategic avoidance
(\emph{IN02} \mean{4.0}, \emph{PR03} \mean{3.8}).  \emph{Ghostbusters}
maintains competitive scoring through creature capture (\emph{IN02}
\mean{4.2}, \emph{PR03} \mean{3.8}).

\subsubsection{Gender-inclusive Projects}

Three projects stand out with the lowest stereotype scores, thus
demonstrating how programming learning can be achieved without gender
bias according to our raters. The absence of anthropomorphic
characters in all of those projects may eliminate opportunities for
gender stereotyping while maintaining interactive discovery and
creative expression.
\emph{Paint Box} (ID 63473366, \score{1.6}) achieves the lowest
stereotype score amongst all tutorials by combining creative
expression with computational thinking (variable manipulation) without
relying on gendered characters or themes.
\emph{Musical Droplets} (ID 1111576868, \score{1.6}) similarly
balances artistic and technical elements through interactive sound
generation.
\emph{Make a Mouse Trail} (ID 1105118803, \score{1.7}) focuses on
visual programming mechanics where users control a star sprite to
create mouse trails. The project emphasises the logic behind visual
effects rather than character representation or narrative themes.

\interpretation{}{One fifth of the \Scratch tutorials contain gender
  stereotypes: Projects targeting girls emphasise appearance and
  social interaction using simpler programming constructs; projects
  targeting boys focus on competition and game mechanics and are more
complex.}

\subsection{RQ2: LLM-based Bias Detection}\label{sec:rq2}

\begin{figure}[t]
  \includegraphics[width=\linewidth]{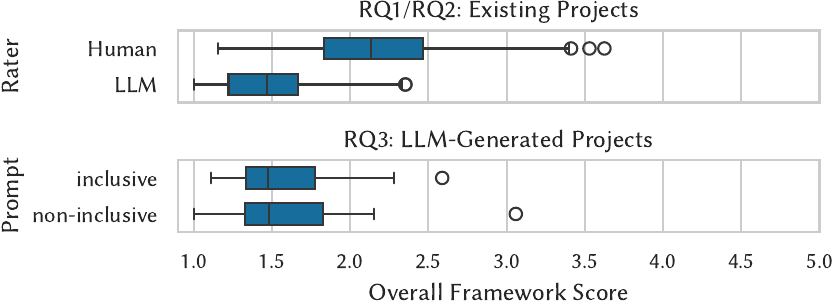}
  \caption{Overall framework scores for existing and generated \Scratch tutorials. Both prompt variants in RQ3 were evaluated by human raters.\label{fig:rq1-rq2-rq3-comparison}}
\end{figure}

\begin{figure}[t]
\includegraphics[width=\linewidth]{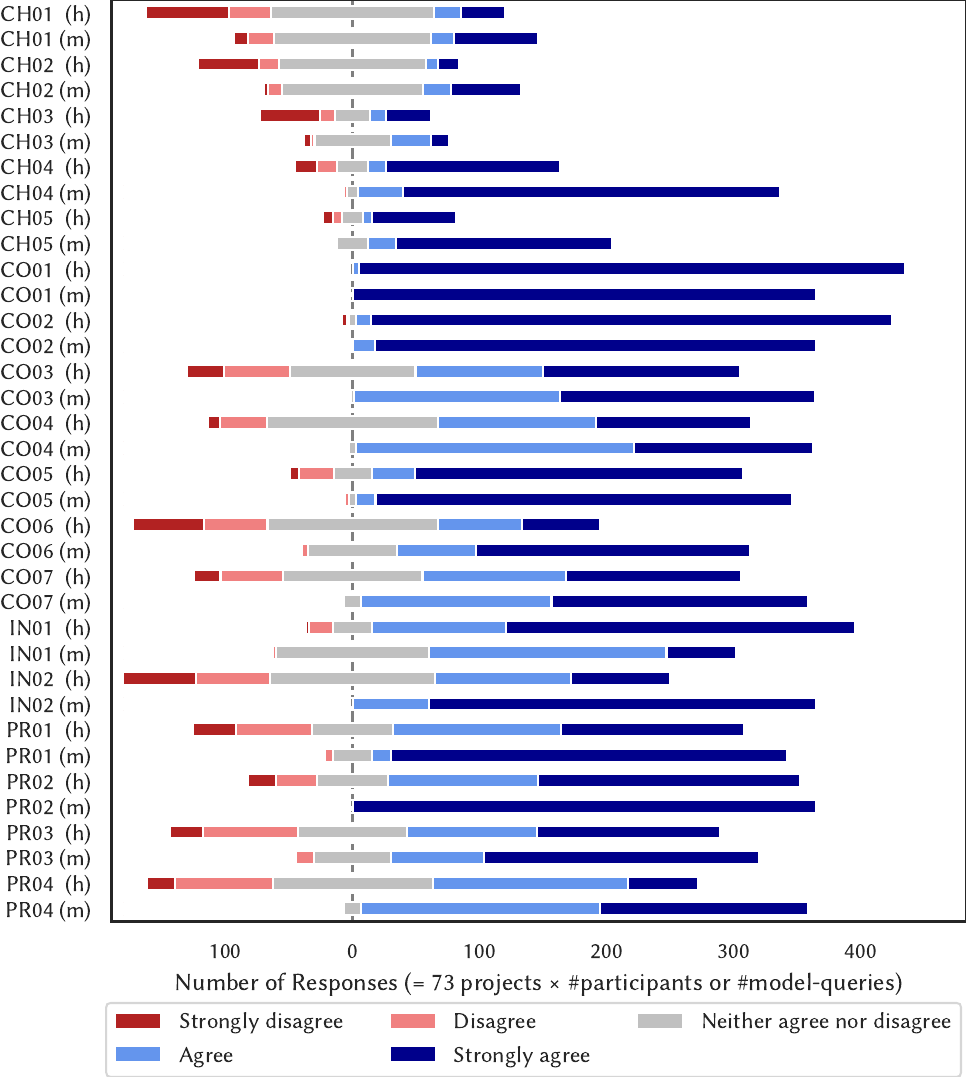}
\caption{Comparison between human~(h) and \llm~(m) ratings for
    existing \Scratch projects across framework
  categories.\label{fig:rq1-rq2-likert}}
\end{figure}

Overall, when using the prompt variant without the framework, the \llm
classified only two out of the 73 projects as not being
gender-inclusive (2.74\,\%), with one targeting girls and one
targeting boys. Using the second prompt variant including the
framework, all projects were classified as inclusive. Thus, the \llm
appears to be unable to properly detect stereotype smells.
\Cref{fig:rq1-rq2-rq3-comparison} compares the overall framework
scores between human raters and the \llm, confirming substantial
disagreement: The mean score of human ratings is 2.17 while the mean
score of \llm ratings is 1.48. The difference between ratings is
statistically significant with \mbox{$p < 0.001$} according to a
Mann-Whitney U-test~\cite{Mann1947}.
\Cref{fig:rq1-rq2-likert} shows strong disagreement between human and
\llm ratings for some framework criteria, and others with only small
differences~(\emph{CH}, \emph{CO01}, \emph{CO02}).

For the \emph{Memory} tutorial (ID 34874510, \mscore{2.1},
\hscore{2.8}) the human raters and the \llm agree that it is targeted
at girls. The \llm detected the same items as the human evaluators in
the project as containing high stereotype smells, although its scores
were slightly lower compared to the human evaluation: \emph{CH01}
(\mmean{4.8}, \hmean{5.0}), \emph{CH02} (\mmean{4.0}, \hmean{4.6}),
\emph{CH04} (\mmean{3.4}, \hmean{4.3}). \Cref{fig:rq1-rq2-likert}
generally shows agreement between human and \llm ratings across these
categories.

Naturally, with the \llm classifying most projects as inclusive, there
are many projects on which both the model and human raters agree. For
example, \emph{Art Alive} (ID 1106198418, \mscore{1.0}, \hscore{1.9})
has the lowest score indicated by the \llm. It represents an
interactive animation where you can click on abstract and colourful
people which are moving or changing their appearance.

Even though the \emph{Dress Up Tera}
project~(cf.~\cref{fig:example:girl}) was ranked as the most
girl-stereotypical one by human raters, it was judged by the \llm to
be inclusive (\mscore{1.3}, \hscore{2.9}). Its main reasoning in both
prompt variants is that the non-human character has no male or female
coded features~(e.g., eyelashes), and that the clothing options are
neutral.
Accordingly, the \llm's stereotype smells compared to human evaluators
were lower in character categories (\mbox{\emph{CH01--04}} ranging
from \mmean{\text{1.0 to 3.0}}, \hmean{\text{4.0 to 4.5}}), content
(\emph{CO03} \mmean{1.4}, \hmean{4.2}; \emph{CO07} \mmean{1.4},
\hmean{4.3}), and the balance of creativity and computational elements
(\emph{PR04} \mmean{1.2}, \hmean{3.3}). \Cref{fig:rq1-rq2-likert}
generally shows stronger disagreement between human and \llm ratings
across the latter three categories.
Similarly, two of the projects containing the least stereotype smells
as identified by the \llm as gender-inclusive, were classified by the
humans as targeted for the girls: \emph{Mandala} (ID 536953224,
\mscore{1.0}, \hscore{2.3}) and \emph{Flower Generator} (ID 253355932,
\mscore{1.1}, \hscore{2.5}).
The highest-ranked stereotype-smelling project for boys as identified
by human evaluators, \emph{Clone Wars}, was identified by the \llm to
be gender-neutral (\mscore{2.1}, \hscore{2.8}; \emph{CH01}
\mmean{3.0}, \hmean{3.5}; \emph{IN02} \mmean{1.0}, \mmean{4.0}).
Interestingly, the \llm classified the project clearly as a game with
an even higher stereotype score for game-related content than the
human evaluators (\emph{PR03} \mmean{4.2}, \hmean{3.8}).

On the other hand, even though human evaluators classified the
\emph{Find the Clone} (ID 259020474, \mscore{2.3}, \hscore{2.0})
project as gender-inclusive, the \llm identified it as targeted
towards boys; in fact this is the tutorial with the highest number of
stereotype smells reported by the \llm.
In this project, the character is randomly selected out of a diverse
set of characters, which may have made it difficult for the \llm to
detect the gender-neutral nature of the task. The project is about
finding this particular character amongst different characters,
similar to the game \emph{Where's Waldo}. The \llm assigned stereotype
scores to the character categories: \emph{CH01} (\mmean{4.6},
\hmean{1.7}), \emph{CH02} (\mmean{3.5}, \hmean{2.3}), \emph{CH03}
(\mmean{4.6}, \hmean{2.8}).

\interpretation{}{The results indicate that current \llms are not
  effective at detecting gender stereotype smells in \Scratch
tutorials.}

\subsection{RQ3: Bias in LLM-Generated Tutorials}

\begin{table}[t]
  \caption{Overall verdict of gender-inclusiveness and usability~(potentially after \emph{mod}ification) of the generated projects.\label{tab:gen-prog-verdict}}
  \footnotesize \setlength{\tabcolsep}{2pt}
  \begin{tabular}{l rrr c rrrr}
    \toprule Category & \multicolumn{3}{c}{Inclusiveness} & & \multicolumn{4}{c}{Use in Classroom}\\
    \cmidrule(r){2-4}\cmidrule(l){6-9}
    i=inclusive & boy & girl & incl. & & as-is & some mod. & lots of mod. & not\\
    \midrule Animation       & 0 & 0 & \textbf{6} && 0 & \textbf{6} & 0 & 0\\
    Animation (i)   & 0 & 0 & \textbf{6} && \textbf{6} & 0 & 0 & 0\\
    Community       & 0 & 0 & \textbf{6} && 0 & \textbf{6} & 0 & 0\\
    Community (i)   & 0 & 0 & \textbf{6} && 0 & \textbf{6} & 0 & 0\\
    Extensions      & 0 & 0 & \textbf{6} && \textbf{5} & 0 & 0 & 1\\
    Extensions (i)  & 0 & 0 & \textbf{6} && \textbf{4} & 1 & 0 & 1\\
    Games           & 0 & 0 & \textbf{6} && \textbf{5} & 1 & 0 & 0\\
    Games (i)       & 0 & 0 & \textbf{6} && 2 & \textbf{4} & 0 & 0\\
    Art             & 0 & 1 & \textbf{5} && 0 & \textbf{6} & 0 & 0\\
    Art (i)         & 0 & 0 & \textbf{6} && 0 & \textbf{5} & 1 & 0\\
    Maths           & 1 & 0 & \textbf{4} && \textbf{3} & \textbf{3} & 0 & 0\\
    Maths (i)       & 1 & 1 & \textbf{4} && 0 & \textbf{5} & 1 & 0\\
    Music           & 0 & 0 & \textbf{6} && \textbf{6} & 0 & 0 & 0\\
    Music (i)       & 0 & 0 & \textbf{6} && 1 & \textbf{5} & 0 & 0\\
    Stories         & 0 & 0 & \textbf{6} && 1 & \textbf{5} & 0 & 0\\
    Stories (i)     & 0 & 0 & \textbf{6} && \textbf{6} & 0 & 0 & 0\\
    \bottomrule \end{tabular}
\end{table}

\Cref{tab:gen-prog-verdict} compares the results of non-inclusive and
inclusive prompts across each of the eight project categories. Most
educators judged the generated projects to be gender-inclusive across
all project categories and both prompt types. Only for the \emph{Art}
and \emph{Maths \& Science} project categories some projects were
judged to be gender-stereotypic.
Most educators also stated that they would use the generated projects
to teach programming to children with no or only little modification.
\Cref{fig:rq1-rq2-rq3-comparison} compares the overall framework
scores between the two prompt variants.
The low scores indicate few stereotype smells for both variants, but
the difference between prompt designs is minimal and not statistically
significant~(Mann-Whitney U\@: $p=0.89$).
All detailed \llm-generated project descriptions and the educator
ratings can be found in our replication
package~\cite{KoliReplicationPackage}.

\subsubsection{Animation}

Both versions demonstrate comparable scores across almost all
categories, with the non-inclusive version showing slightly higher
presence of stereotype smells (inclusive \iscore{1.3}, non-inclusive
\nscore{1.5}).
However, the non-inclusive version received a lower rating for skill
diversity (\emph{IN01}). While the inclusive variant contains a
storytelling aspect as an animated story with collaborative elements,
the non-inclusive version features a character \emph{only} dancing on
a stage, offering fewer skill development opportunities.

\subsubsection{Community and Kindness}

The non-inclusive version contains slightly more stereotype
smells~(\iscore{1.6}, \nscore{1.9}).
Higher scores for the non-inclusive version are particularly evident
in the character categories (\emph{CH01–-03}), which might be
attributed to the suggested sprite \emph{Avery}, a female black child
who sits on a park bench and smiles when the user picks up litter.
In the inclusive version, characters appear more stereotyped by
occupational roles (\emph{CH04}) and show more emotionally
stereotypical attributes (\emph{CH05}). In particular, the inclusive
community project presents more stereotype smells regarding the
reflection of diverse interests beyond stereotypical preferences, with
almost double the score (\emph{CO07}, \imean{3.0}, \nmean{1.6}). This
difference may stem from the girl (Ellie) gardening and happily
watering dry flowers, whilst Sam, the boy, organises a book exchange,
alongside a dog and a robot who help with recycling and technology.
An advantage of the inclusive project is that it offers more different
kinds of skills (\emph{IN03}, \imean{1.8}) compared to the
non-inclusive version (\nmean{3.3}), which may be due to storytelling
elements and the interactive style between the different characters.

\subsubsection{Extensions}

Overall, both versions have comparable levels of gender stereotype
smells across almost all categories (both \score{1.7}).
While most educators would use either variant project as presented,
one would not use either of them at all. The non-inclusive program is
similar to the one shown in \cref{fig:example:girl}, but replaces the
character with a robot. The educator specifically noted that this mix
of girl- and boy-stereotyped content may result in the project
appealing to neither rather than making it inclusive.
Educators identified more stereotype smells regarding the balance of
competitiveness and cooperation (\emph{CO06}) in the inclusive
project, as it involves an invention show. In contrast, the
non-inclusive version features only one sprite for decoration. This
difference may also contribute to educators' perception that the
inclusive version appears less adaptable in terms of student
preferences, given that the \emph{Decorate your Robot} project offers
greater sprite customisation opportunities~(\emph{IN02}).

\subsubsection{Games}

The inclusive project (\iscore{1.4}) shows fewer stereotype smells
than the other version (\nscore{1.7}).
This difference is particularly evident in the categories about
stereotyped characters by occupational roles~(\emph{CH04}), diverse
applications beyond games and balance of creative and artistic
programming (\emph{PR03, PR04}), and addressing different kinds of
skills~(\emph{IN01}).
However, the inclusive version also presents higher stereotype
indicators in some categories, such as balance of competitiveness
versus cooperation (\emph{CO06}) and themes reflecting diverse
interests beyond stereotypical preferences~(\emph{CO07}).

\subsubsection{Interactive Art}

The inclusive version (\iscore{1.7}) shows more stereotype smells than
the non-inclusive version (\nscore{1.3}).
The non-inclusive variant of the program allows the user to draw
colourful shapes, e.g., stars, on the screen using the mouse. While
most educators agreed that this project is gender-inclusive, one
considered it to be targeted at girls.
This difference is particularly evident in the character categories
(\emph{CH}). In the non-inclusive version, the drawing sprite is a
simple shape, e.g., a star. In contrast, the inclusive version
features \emph{Ada}, a brightly dressed girl with a paintbrush,
\emph{Jay}, a boy in a green shirt and trainers holding a spray can,
and \emph{Quark}, a friendly robot with colourful arms and a smile.
The pairing of Ada with traditional painting and Jay with
spray-painting (usually associated with graffiti) may contribute to
stereotypical associations.

\subsubsection{Maths and Science}

Both versions are comparable in terms of stereotype
smells~(\iscore{1.7}, \nscore{1.7}), yet the inclusive version
indicates slightly more stereotype smells.
One educator deemed the non-inclusive version to be targeted at boys.
In the program the player answers maths questions by catching the
correct answer option falling down across the screen.
For the inclusive version, opinions differ: four educators rate it as
gender-inclusive, while one rates it as targeted for girls and one for
boys.
The inclusive project shows more stereotype smells regarding balance
of competitiveness versus cooperation (\emph{CO06}) and  themes
reflecting diverse interests beyond stereotypical
preferences~(\emph{CO07}).
However, the inclusive version performs better with fewer stereotype
indicators in character categories (\emph{CH01, CH02, CH03, CH05}),
addressing different kinds of skills (\emph{IN01}), and presenting
programming as having applications beyond games and competition
(\emph{PR03}).

\subsubsection{Music}

Slight differences are observed between versions, with the inclusive
one having more stereotype smells (\iscore{1.6}) than the
non-inclusive version (\nscore{1.3}).
The introduction of multiple \enquote{diverse} characters in the
inclusive version may create opportunities for stereotypical
associations that are absent in the other variant in which students
create their own abstract musical instruments.

\subsubsection{Stories}

Both versions contain similar amounts of stereotype
smells~(\iscore{1.7}, \nscore{1.5}), but distributed across different
categories.
The inclusive project shows more stereotype indicators regarding the
content, style, and visual interface (\emph{CO03} \imean{2.5},
\nmean{1.3}; \emph{CO04} \imean{2.3}, \nmean{1.2}).
Both projects follow a \enquote{choose your adventure} theme. While
the inclusive variant focuses on a pirate-themed lost treasure, the
other project leaves a choice of adventure~(e.g., forest or beach).

\interpretation{}{\llms are able to generate projects with relatively
  few gender stereotype smells that are ready to use in class.
However, not even gender-inclusive prompting guarantees bias free
tutorials.}

\section{Discussion}

\subsection{Applicability of LLMs} 

Our findings reveal a contrast between \llm-generated tutorials and
existing educational materials in presence of stereotype smells.
Educator ratings up to 5 across several categories suggest existing
tutorials contain more obvious and pronounced stereotype smells than
generated ones with means below 3.3.
This suggests that \llm-generated content contains more subtle forms
of stereotyping compared to traditional educational resources,
confirming recent work showing that \llms create personas that appear
sophisticated but remain rather narrow regarding their
diversity~\cite{venkit2025}.
However, this apparent improvement comes with a tradeoff: the
stereotypes in generated content are more nuanced and potentially
harder to detect. The subtlety of these biases could influence young
learners without explicit recognition by educators or students
themselves. Further and refined guidelines could help both educators
and \llms.

The difference in performance of \llms for generating vs. evaluating
projects might also be a result of the \llm training data.
We hypothesise that the \llm has seen a lot of general rather than
programming specific training data about gender-inclusiveness.  When
generating new projects~(RQ3), it can utilise this information for
natural language project descriptions that are gender-inclusive.
However, when judging the inclusiveness of existing
tutorials, it needs to apply these general concepts in the specific
context of \Scratch programs. RQ2 suggests the \llm 
focuses on the characters rather than the program’s
content~(cf.~\cref{sec:rq2}).
In combination with the higher agreement between human and model
raters for the character categories~(\emph{CH},
cf.~\cref{fig:rq1-rq2-likert}), we see that the \llm successfully
recognises the figures in the images and considers them in its
decision. However, it might be difficult for the model to understand
the objective of the executed program, which is relevant to detect
stereotyped behaviour of the characters~(\emph{CO02}, \emph{CO06}).
Since the \Scratch forums contain mostly code snippets rather than
complete programs, there is insufficient task-related \llm training
data. This calls for research on improving \llms for handling \Scratch
code.

\subsection{Paradox of Inclusive Character Design}

Across our analysis we saw that adding diverse characters may
introduce new stereotypes while attempting to be inclusive. In
particular, in RQ3 we observed that the intention to represent
diversity leads to character combinations that reinforce binary
thinking.
This reflects broader challenges in \llm training data, which
predominantly contains binary gender concepts~\cite{kotek2023},
despite the potential for \llms to serve as evaluation
tool~\cite{ye2024}.
Our gender-inclusive prompting mostly resulted in projects featuring
one girl and one boy working together, typically accompanied by a
\enquote{neutral} character such as a \enquote{friendly robot} or a
parrot. This reveals several problematic assumptions:
Firstly, it reinforces binary gender concepts by consistently pairing
male and female characters while treating everything else as
non-gendered. Secondly, it suggests that \enquote{inclusion} is
achieved simply through numerical representation rather than
meaningful diversity of roles, interests, or capabilities.
Thirdly, although robots are supposedly neutral
characters~(\emph{CO05} is consistently rated better in inclusive
projects), these characters are often implicitly coded as male through
language and behaviour. As one educator noted: \enquote{There
are so many different characters in \Scratch, I do not understand why
it so often has to be a \enquote{friendly robot} as helper figure.}
Future qualitative research should explore how educators would adjust
tutorials and establish a mutual understanding of what gender
inclusion in tutorial design means.
Not only the \llm's output, but also our framework itself includes
such binary coding in some criteria~(e.g., \emph{CH01}, \emph{CH02},
\emph{CO03}). Future refinements should aim to better accommodate
non-binary perspectives, moving beyond the current binary-plus-neutral
model.

\subsection{Challenging Gender Norms}

The stereotypes found in \Scratch tutorials~(RQ1) and the inability of
the \llm to detect them~(RQ2) reflect wider societal prejudices. For
instance, the overemphasis on appearance-focused or passive characters
in tutorials for girls may discourage them from more technical aspects
of computing.  Similarly, boys face pressure to conform to
traditionally masculine traits reinforced by games, such as
competitiveness or aggression, which may limit their emotional
expression or interest in non-competitive, creative projects.

To break these patterns, tutorials should subvert traditional gender
roles. For example, male characters could take on roles typically seen
as \emph{feminine}, while female characters could be shown solving
complex technical problems, challenging the idea that girls are not
suited for advanced programming tasks. An example is provided by the
\llm-generated \emph{Decorate your Robot} project (RQ3) in which a
robot is decorated, which shares similarity with dressing the girl in
\emph{Dress Up Tera} but not its gender bias.

Of course some girls may enjoy appearance-focused activities as much
as games~\cite{spieler2019}, and some boys may prefer creative over
competitive projects. However, the concern lies in systematic
overrepresentation that may cause implicit barriers for learners whose
interests do not align with stereotypical expectations. Thus, the goal
is expanding rather than eliminating representation.
This is particularly important considering that our findings confirm
observations made more than two decades ago by Bhargava et
al.~\cite{bhargava2002}, who reported bias in the program
\enquote{Dream Doll Designer} targeted at girls. This demonstrates
that appearance-focused, stereotypical content has long persisted in
educational programming environments.

 \section{Conclusions and Future Work}

The \Scratch gender bias evaluation framework proposed in this paper
provides a means to assess \Scratch programming tutorials with respect
to stereotype smells. While \llms at this point do not yet seem
capable of properly \emph{identifying} these stereotype smells, they
have nevertheless demonstrated potential in helping educators
\emph{create} programming tutorials free of gender bias. Given the
framework and this insight opens up many avenues for further research
on reducing gender bias in early programming education. On the
technical side, since the \llm is able to come up with inclusive ideas
future work could explore the use of \llms to improve existing
tutorials, for example by suggesting fixes for
stereotype smells. As we observed \llms to be more consistent than
human raters, they could also help during future iterations of
improving our proposed gender bias evaluation framework. However,
the limitations of \llms in creating and handling \Scratch code will
need to be addressed as well. Furthermore, evaluations of the
real-world impact and student responses to stereotype smells are
equally important.

\balance \bibliographystyle{ACM-Reference-Format}
\bibliography{koli-llms-bias-scratch-gender,bib2}

\end{document}